# Stable Néel-Twisted Skyrmion Bags in a van der Waals Magnet Fe$_{3-x}$GaTe$_2$ at Room Temperature

*Jialiang Jiang[1#], Yaodong Wu[2,3#*], Lingyao Kong[1], Yongsen Zhang[1], Sheng Qiu[3], Huanhuan Zhang[1], Yajiao Ke[4*], Shouguo Wang[5], Mingliang Tian[1,3], and Jin Tang[1,3*]*

[1]School of Physics and Optoelectronic Engineering Science, Anhui University, Hefei, 230601, China

[2]School of Physics and Materials Engineering, Hefei Normal University, Hefei 230601, China

[3]Anhui Province Key Laboratory of Low-Energy Quantum Materials and Devices, High Magnetic Field Laboratory, HFIPS, Chinese Academy of Sciences, Hefei 230031, China

[4]School of Physics and Mechanics, Wuhan University of Technology, Wuhan 430070, China

[5]Anhui Provincial Key Laboratory of Magnetic Functional Materials and Devices, School of Materials Science and Engineering, Anhui University, Hefei 230601, China

*Corresponding author: wuyaodong@hfnu.edu.cn; keyajiao@whut.edu.cn; jintang@ahu.edu.cn

[#]These authors contributed equally.

# Abstract

Magnetic skyrmion bags with diverse topological charges $Q$, offer prospects for future spintronic devices based on freedom of $Q$. While their emergence in van der Waals magnets holds the potential in developing $Q$-based 2D topological spintronics. However, previous room-temperature skyrmion bags necessitate special anisotropy engineering through disorder Fe intercalation, and the stable phase diagram for skyrmion bags across room temperature regions is lacking. Here, we demonstrate the observation and electrical manipulation of room temperature skyrmion bags in $Fe_{3-x}GaTe_2$ without specially designed Fe intercalation. Combining the pulsed currents with the assistance of magnetic fields, skyrmion bags with various topological charges are generated and annihilated. Especially double nested skyrmion bags are also discovered at room temperature. The stable temperature-field diagram of skyrmion bags has been established. We also demonstrate the electrical-controlled topological phase transformations of skyrmion bags. Our results will provide novel insights for the design of 2D skyrmion-based high-performance devices.

Magnetic topological charge $Q$ is an important index characterizing the fundamental properties of topological spin textures, such as magnetic skyrmions that own an integer topological charge $Q$. The fascinating electromagnetism of skyrmions, including ultralow driving current density, topological stability, skyrmion hall effect, and novel electrical transport behaviors, is closely related to $Q$.[1-13] Topological spin textures with multiple $Q$ hold promise for building high-performance spintronics. Moreover, the diversity of topological objects could also come from the Hopf indices, which could also hold potential for data storage applications associated to integer-valued topological invariants.[14-16] Recently, skyrmion bags (also known as skyrmion sacks)[17-19], i.e. skyrmions encircled by closure spirals, have been theoretically proposed and shown to possess arbitrary values of $Q$.[18-24] For instance, a skyrmionium consists of two nested skyrmions with opposing topological charges, resulting in a net charge ($Q$) of 0. In 3D chiral magnets, skyrmion bags are extended to be skyrmion bundles, whose surface twists turn to high-order antiskyrmions.[25, 26] Notably, these aforementioned spin textures also can be categorized as composite skyrmions.[17]

Skyrmion bundles were first presented by reversing the magnetic field in chiral FeGe lamella.[25] Subsequently, skyrmion bundles that stabilize at room temperature with zero magnetic fields have been discovered in $Co_8Zn_{10}Mn_2$, contributing to practical $Q$-based topological spintronic applications.[26] Besides, dipolar skyrmion-bubble bags existing in X-type $Sr_2Co_2Fe_{28}O_{46}$ hexaferrite have been demonstrated experimentally, enriching the family of magnetic solitons with multi-$Q$ characteristics.[27] It is worth noting that the dipolar skyrmions in Co/Ni multilayers consist of a single closed domain wall with various numbers of Bloch and Néel segments, presenting an alternative approach for stabilizing skyrmions with arbitrary $Q$.[28]

Recently, two-dimensional (2D) van der Waals (vdW) magnets have demonstrated impressive exfoliation performance, abundant topological phenomena, and novel device architectures.[29-35] The distinguished performance of 2D vdW magnets prompts them to apply to multifunctional spintronic devices, including magnetic tunnel junctions, spin valves, spin field-effect transistors, etc.[36-38] Skyrmion bags emerging in 2D vdW magnets offer exciting opportunities for 2D multi-$Q$ spintronic device applications.[17, 39, 40] In contrast to Bloch-twisted skyrmion bags in chiral $B_{20}$ magnets and dipolar uniaxial magnets, skyrmion bags in 2D magnets exhibit Néel-twisted spin twists, which is attributed to the interfacial Dzyaloshinskii–Moriya interaction (DMI) induced by dislocated atoms in the 2D layer.[41-43] However, room-temperature skyrmion bags reported in 2D vdW magnets necessitate special anisotropy engineering through disorder Fe intercalation,[40] and the stable phase diagram of skyrmion bags as a function of temperature and magnetic field in 2D magnets across room temperature regions is established not yet.

In this work, we experimentally present the observation and electrical manipulation of room temperature skyrmion bags in $Fe_{3-x}GaTe_2$ thin lamellas using Lorentz transmission electron microscopy (Lorentz TEM). The $Fe_{3-x}GaTe_2$ lamellas used in our experiments do not involve specially designed Fe intercalation. By employing pulsed currents and magnetic fields, diverse skyrmion bags emerge within a specific magnetic field range, transforming into low-$Q$ skyrmion bags as the fields increase. Notably, double nested skyrmion bags at room temperature are also observed, which appear more complex spin textures. The stable diagram of room temperature skyrmion bags with zero $Q$ ($Q = 0$ bag) in dependence of temperature $T$ and field $B$ has been summarized. Additionally, we investigate the dynamic response of the skyrmion bags to pulsed currents. Micromagnetic simulations have been conducted to illustrate the detailed spin arrangement within the skyrmion bags. Our findings are expected to enhance the understanding of topological spin textures and promote the advancement of applications for 2D multi-$Q$ devices.

$Fe_{3-x}GaTe_2$ is a newly discovered 2D vdW ferromagnet known for having the highest $T_c$ (340-380 K) among 2D magnets, along with strong uniaxial magnetic anisotropy.[41, 44-52] It has a centrosymmetric crystal structure of $P6_3/mmc$ space group, typically indicative of the absence of the Dzyaloshinskii–Moriya interaction (DMI). Recent research indicates that local Fe deficiencies cause a slight displacement of Fe atoms along the $c$-axis, which breaks the centrosymmetry and induces the interfacial DMI across the material, resulting in the stabilization of Néel-type skyrmions.[40, 41] As shown in Figure S1a, Fe-Ga layers are sandwiched by Te layers along the $c$-axis of the crystal. While in the crystallographic $ab$-plane of $Fe_{3-x}GaTe_2$, the schematic arrangement of the atoms presents a hexagonal lattice structure (Figure S1b). High-resolution transmission electron microscopy (HRTEM) image and the corresponding selected area diffraction (SAED) of the $Fe_{3-x}GaTe_2$ thin lamella with a crystallographic (001) plane confirm the good quality of the sample (Figure S2). Figures S1c and 1d display the energy dispersive X-ray spectroscopy (EDS) analysis of the $Fe_{3-x}GaTe_2$ crystal surface, where three composite elements distribute uniformly. The atomic ratio of iron, gallium, and tellurium is 2.89: 1: 2.03, indicating the possible existence of iron vacancies. Macroscopic magnetic measurement is conducted utilizing the magnetic properties measurement system (MPMS). Figure S1e shows the temperature-dependent magnetization of $Fe_{3-x}GaTe_2$, the significant difference between the curves of in-plane ($B//ab$-plane) and out-of-plane ($B//c$ axis) orientations suggesting a perpendicular magnetic anisotropy. Besides, both the field-cooled (FC) and zero-field-cooled (ZFC) curves are presented in Figure S1e as solid and dashed lines. By calculating the derivative of the magnetization curves along the $c$-axis, the $T_c$ of $Fe_{3-x}GaTe_2$ single crystal in this work is approximately 344 K, much higher than most vdW ferromagnets. Figure S1f and Figure S3 show the magnetic field-dependent magnetization at different temperatures, where $B$ is parallel or perpendicular to the $c$-axis, respectively. The isothermal magnetization curves further verify the perpendicular magnetic anisotropy of $Fe_{3-x}GaTe_2$.

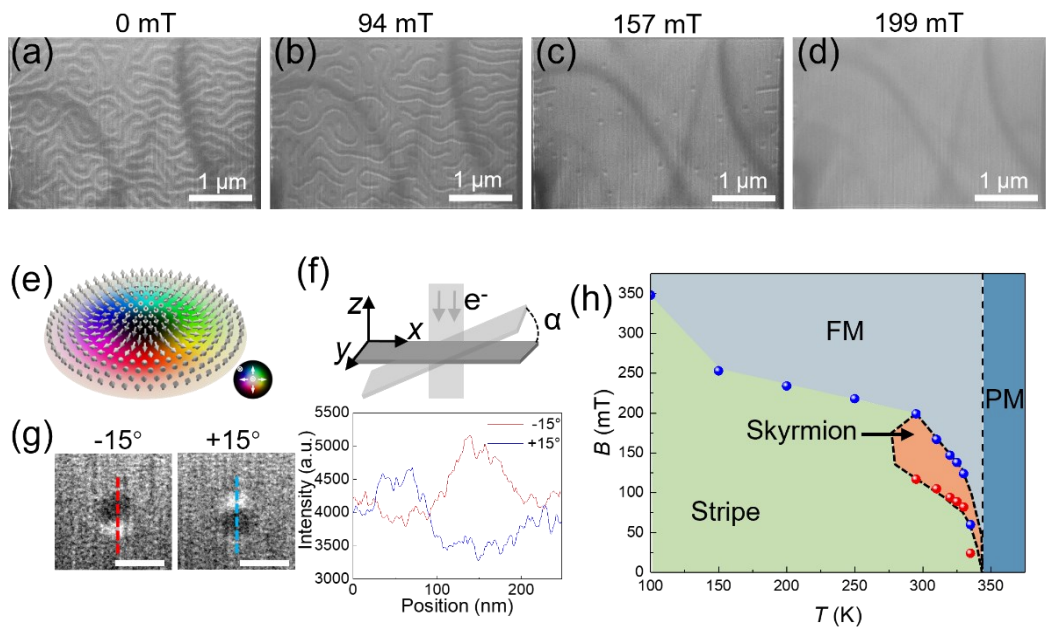

**Figure 1.** Field-driven evolutions of the Fe$_{3-x}$GaTe$_2$ thin lamella at $T = 295$ K. (a)-(d) representative Fresnel images recorded in a $B$-increasing process at $B = 0$, 94, 157, and 199 mT, respectively. The defocused distance is −2.8 mm. Scale bars, 1 μm. The sample is tilted at −15°. (e) Simulated schematic diagram of a single Néel-type skyrmion. The color wheel indicates the orientation of magnetization. Dark and white contrasts represent out-of-plane down and up magnetizations, respectively. (f) Schematic diagram of the incident electron beam transversing the tilted sample plane at an angle $\alpha$. (g) Line scans of the single Néel skyrmion at −15° (red line) and 15° (blue line). Scale bars, 200 nm. (h) Experimental phase diagram of the magnetic textures in Fe$_{3-x}$GaTe$_2$ thin lamella as a function of magnetic field $B$ and temperature $T$.

We fabricated thin lamellas (~170 nm) from the *ab*-plane of single Fe$_{3-x}$GaTe$_2$ crystals using focus ion beam (FIB) technology, followed by an investigation of the static magnetic evolutions of the samples at 295 K. Figures 1a-d show the Fresnel images recorded by Lorentz TEM with an out of plane magnetic field ($B$//$c$-axis). At zero field, stripe domains are observed as the stable ground magnetic state (Figure 1a), which then become sparse and narrow as the field increases (Figure 1b). Figure 1c shows that, at $B = 157$ mT, all stripe domains annihilate, transforming into Néel-type skyrmions ($Q = -1$) before gradually diminishing as the field continues to rise. Finally, when the field reaches about 199 mT, the lamella is saturated with a uniform Fresnel contrast. The emergence of Néel-type skyrmions in Fe$_{3-x}$GaTe$_2$ is attributed to the interfacial DMI resulting from iron vacancies,[41, 50] as evidenced in the EDS analysis (Figure S1d). Figure 1e shows the schematic structure of a simulated Néel-type skyrmion. When the lamella is perpendicular to the electron beam, there is no contrast in the Fresnel images due to mutual cancellation of the electron beam inflection. When the sample tilts to a positive or negative degree, the Néel-type skyrmions with half-bright and half-dark Fresnel contrast can be observed, reversible based on the tilt angle (Figures 1f and 1g). The simulated Fresnel contrast of Néel-type skyrmion closely matches the experimental results (Figure S4). Figure 1h summarizes the magnetic phase diagram of the magnetic spin textures as a function of the magnetic field $B$ and temperature $T$. The skyrmions are only present near $T_c$, while at low temperatures, no skyrmion is observed under the perpendicular fields.

Next, we will discuss the generation process of the Néel-twisted skyrmion bags in $Fe_{3-x}GaTe_2$. Initially, the magnetic field is set at about −8 mT, stabilizing stripe domains within the sample at 295 K (Figure 2a). Subsequently, a pulsed current with a width of 40 ns and a density $j$ of ~ 4.11 × $10^{10}$ A m$^{-2}$ is applied, resulting in a mixed state consisting of skyrmions with $Q = 1$ and stripe domains (Figure 2b and Movie S1). Figure 2c illustrates the current induced skyrmion count as a function of different current densities, showing a rapid increase in skyrmion count once the $j$ exceeds the critical value $j_c$. Subsequently, by reversing the field to positive values, the mixed state gradually annihilates, transforming into the Néel-twisted skyrmion bags (as shown in the red box) at certain fields (Figures 2d-f).

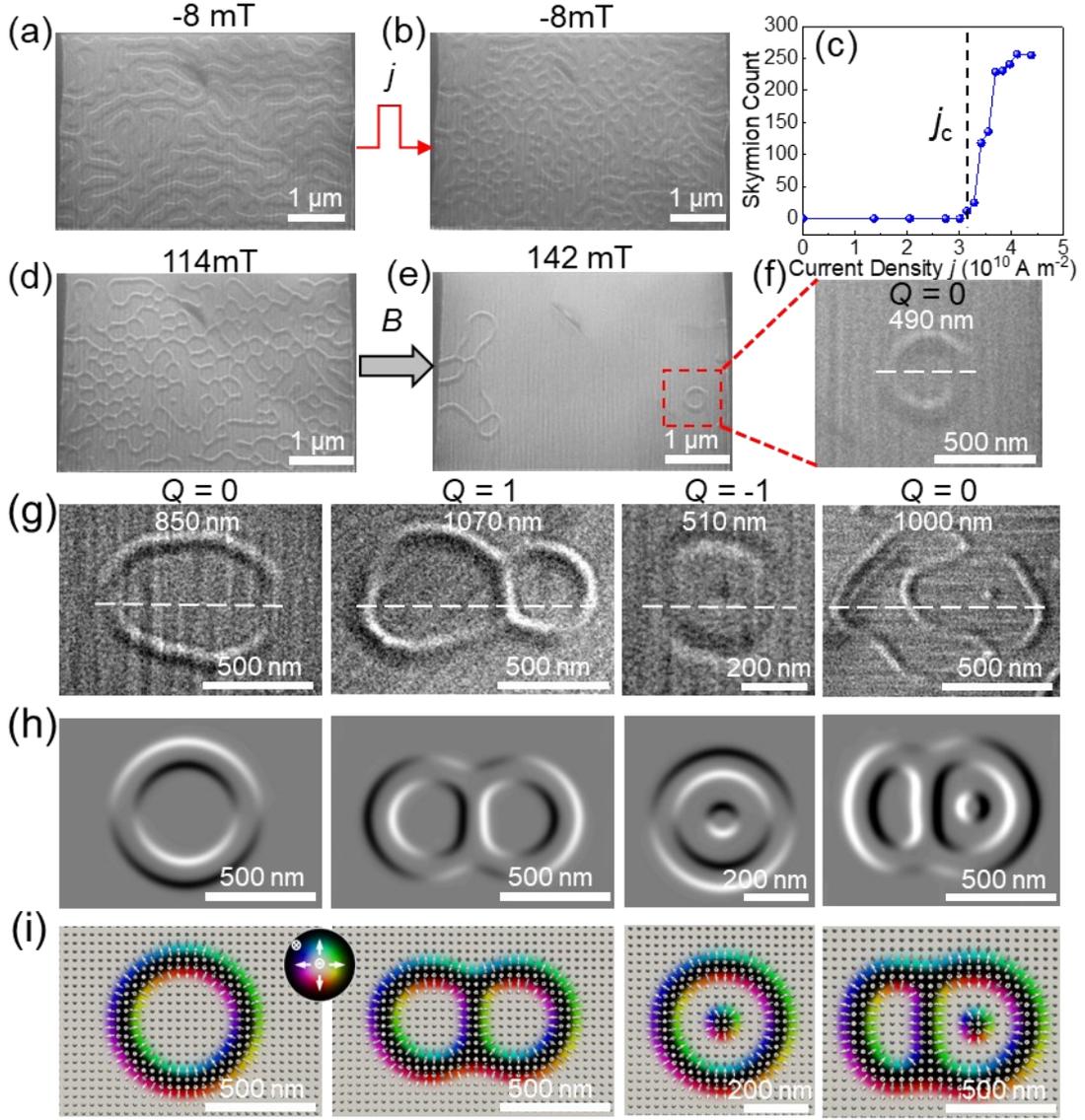

**Figure 2.** Stable Néel-twisted skyrmion bags in Fe$_{3-x}$GaTe$_2$ at room temperature. (a)-(b) Transformation from stripes to mixed skyrmion-stripe states by applying a 40 ns single pulsed current, current density $j \sim 4.11 \times 10^{10}$ A m$^{-2}$. Magnetic field, −8 mT. (c) Skyrmion count generated from stripe domains by single pulsed current as a function of current density $j$, and the critical current density $j_c$ is approximately $3.15 \times 10^{10}$ A m$^{-2}$. The pulse width is 40 ns. $B = -8$ mT. (d)-(e) Generation of skyrmion bags by applying reversed positive magnetic fields. (f) Enlarged image in the red box of (e). $\alpha = +15°$ in (a)-(f). $T = 295$ K in (a)-(f). (g) Experimental Fresnel images of diverse skyrmion bags in Fe$_{3-x}$GaTe$_2$ with $Q = 0, 1, -1$, and 0. The $Q = -1$

bag is observed at −15° and other skyrmion bags in (g) are observed at +15°. $T$ = 295 K. (h) Corresponding simulated Fresnel contrasts of skyrmion bags. (i) Corresponding simulated magnetic configurations of the skyrmion bags. The defocused distance is −2.8 mm. The color wheel indicates the orientation of magnetization. Dark and white contrasts represent out-of-plane down and up magnetizations, respectively.

The skyrmion bag observed in Figure 2f consists of a uniform upwards magnetized region surrounded by a Néel-type domain wall, with a topological charge of zero. Because of its zero topological charge, the $Q = 0$ bag follows a straight trajectory without Hall deflection when propelled by electrical currents. Due to the interfacial DMI in $Fe_{3-x}GaTe_2$, the $Q = 0$ bag remains invisible in Fresnel contrast unless the thin lamella is tilted at an angle from the horizontal plane, revealing the characteristic Néel-twisted features (Figure S5). Diverse skyrmion bags are observed by repeating the generation process multiple times. Figure 2g displays Fresnel images of four different skyrmion bags with topological charge $Q$ = 0, 1, −1, and 0, respectively. Figures 2h-i demonstrate the simulated corresponding spin arrangement and Fresnel contrast of them. Skyrmion bags in the left two columns of Figures 2g-i, consist of inner skyrmions only single nested in the outer skyrmion ring. Each skyrmion inside the bags has $Q = 1$, while the outer skyrmion ring has $Q = -1$, so the skyrmion bags have $Q = N-1$ ($N$ is the inner skyrmion count).[25] The right two columns in Figures 2g-i first demonstrate the double nested skyrmion bags at room temperature, which has been reported both in theory and experiments before but with low temperatures.[17, 19, 22] Additional skyrmions with $Q = -1$ are double nested in the single nested skyrmion bags, so the total $Q$ of them needs to subtract 1, finally equal to −1 and 0. The discovery of double nested skyrmion bags at room temperature will enrich the family of 2D skyrmion bags and promote their practical applications.

The injection of the pulsed current is aimed at creating the initial "seed" state for the formation of skyrmion bags. When a pulsed current with relatively high densities is applied, the temperature of the lamella quickly surpasses its $T_c$, resulting in a swift transition to a uniform ferromagnetic state.[53-55] Subsequently, as the current duration ends, the temperature decreases back to 295 K, causing rapid demagnetization in $Fe_{3-x}GaTe_2$ and leading to the creation of a complex mixed magnetic state. This mixed state is composed of skyrmions and stripe domains. In the subsequent process of field reversal, it breaks down, and annihilates gradually, potentially leading to the automatic appearance of skyrmion bag states. Thus, the state initialized by pulsed currents serves as the "seed" that evolves into skyrmion bags with the assistance of magnetic field $B$. Figure S6 demonstrates the simulated field-driven process for generating skyrmion bags, showcasing close agreement with the experimental results.

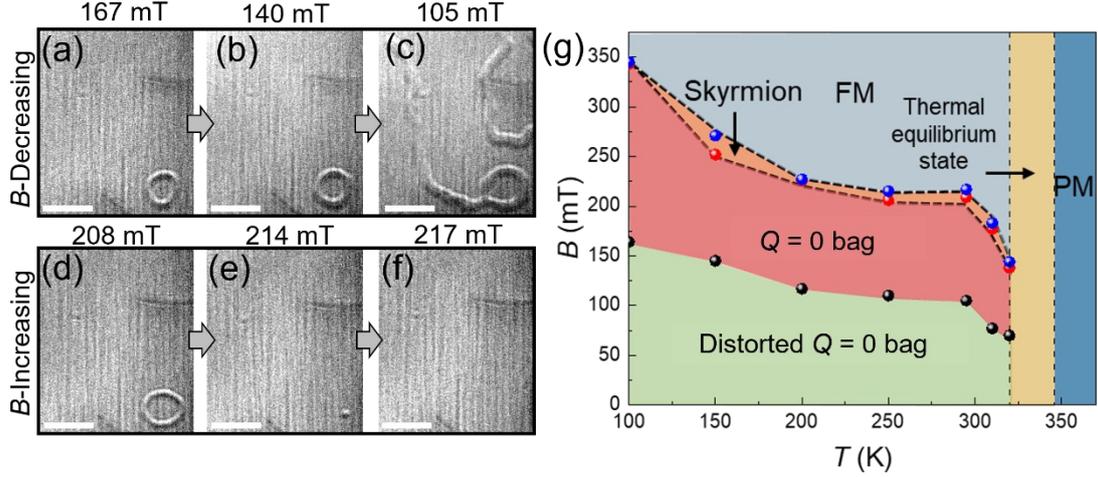

**Figure 3**. Magnetic field-driven evolutions of a single $Q = 0$ bag in $Fe_{3-x}GaTe_2$. (a)-(c) $B$-decreasing process of a single $Q = 0$ bag at 295 K. (d)-(f) $B$-increasing process of a single $Q = 0$ bag at 295 K. The defocused distance is −2.8 mm. Scale bars, 500 nm. $\alpha = +15°$ in (a)-(f). (g) Experimental phase diagram of the $Q = 0$ bag in $Fe_{3-x}GaTe_2$ thin lamella as a function of magnetic field $B$ and temperature $T$.

Next, we investigated the stability of skyrmion bags under varying magnetic fields at different temperatures. A magnetic $Q = 0$ bag, having a zero topological number, is selected as an example to explore its field-driven evolutions, and finally, a phase diagram of the $Q = 0$ bag about temperature $T$ and magnetic field $B$ is established. Figure 3a displays a stable $Q = 0$ bag at $B = 167$ mT with a temperature of 295 K. Decreasing the field causes the $Q = 0$ bag to enlarge and expand its domain wall width slightly, depicted in Figure 3b, where the $Q = 0$ bag size increases at $B = 140$ mT. Further reduction in the field distorts the $Q = 0$ bag, generating stripe domains toward random directions while preserving the domain wall ring feature (Figure 3c). Upon reaching the zero field, the distorted $Q = 0$ bag remains stable with adjacent stripe domains (Figure S7). Once the magnetic field increases, the distorted state will revert to a standard $Q = 0$ bag (Figure S7). As the field $B$ increases, the $Q = 0$ bag contracts, narrowing its domain wall. At $B = 214$ mT, the $Q = 0$ bag suddenly collapses into a single Néel-type skyrmion, leading to annihilation at a saturated field of $B = 217$ mT (Figures 3d-f). Figure S8 illustrates the simulated total energy of the single $Q = 0$ bag as a function of magnetic field $B$. In the field-increasing process, it transforms into a single skyrmion at $B = 250$ mT, accompanied by an obvious energy reduction signifying a shift to a more stable state. Figure S9 shows the field-driven annihilation of a single $Q = 0$ bag at 150 K, suggesting an increasing saturated field compared to 295 K.

Figure 3g demonstrates the magnetic diagram of a single $Q = 0$ bag with a temperature range of 100-320 K, where the $Q = 0$ bag can exist stably. When the temperature is over 320 K, the formation of $Q = 0$ bags is unattainable, leading to a thermal equilibrium state characterized by stable magnetic stripe domains and skyrmions. As shown in Figure S7, the distorted magnetic $Q = 0$ bags recover gradually to their initial states as the field $B$ increases. However, by raising the temperature of the distorted $Q = 0$ bags to 325 K at $B = 0$ mT, the $Q = 0$ bags transform into stripe domains, due to the instability of the metastable $Q = 0$ bag under the impact of thermal fluctuations (Figure S10). In the subsequent $B$-increasing process, no magnetic $Q = 0$ bags emerge from the initial magnetic state (Figure S10).

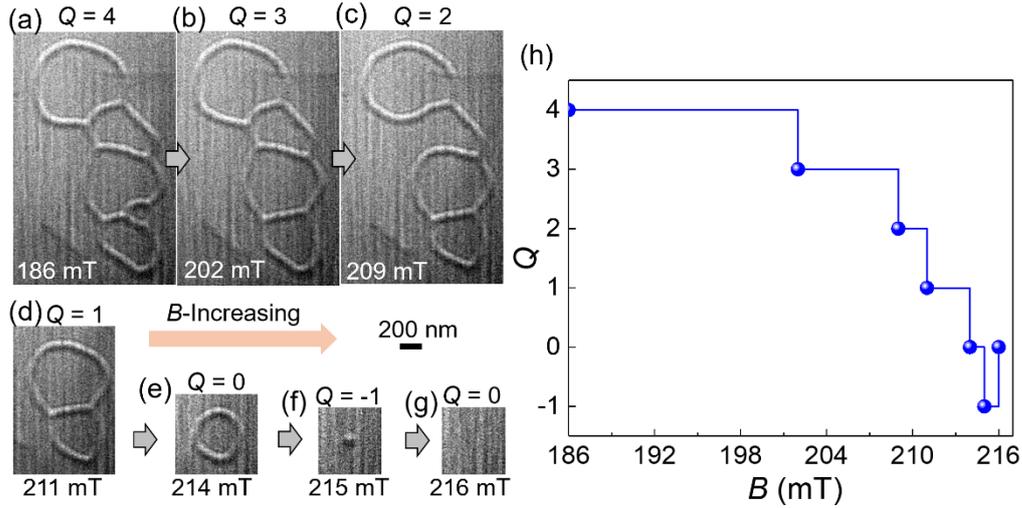

**Figure 4.** Magnetic field-driven evolutions of a skyrmion bag with $Q = 4$ under an increasing perpendicular field at 295 K. (a)-(g) Representative Fresnel images of the initial skyrmion bag with increasing fields of $B = $ 186, 202, 209, 211, 214, 215, and 216 mT. Scale bars, 200 nm. The defocused distance is −2.8 mm. The sample is tilted at +15°. (h) Topological charge $Q$ of the skymion bag as a function of the field $B$.

Furthermore, we also investigate the magnetic evolutions of other skyrmion bags. Figure 4 shows an annihilation process of skyrmion bags with $Q = 4$ at 295 K, where each skyrmion collapses successively with the increasing field, maintaining the overall shape of the structure. Finally, a single skyrmion originates from a $Q = 0$ bag, which subsequently annihilates at a higher field (Figures 4f and 4g). Notably, a slight reduction in the size of the skyrmion bag occurs during the field-driven process until collapses at a significantly high field (Figures 4a-e). This inconsecutive change in size could be attributed to the pinning effect induced by the inhomogeneity and damage in the nanostructured samples. The annihilation sequence varies when a skyrmion bag incorporates a nested skyrmion. As depicted in Figure S11, the nested skyrmion annihilates first as the field increases, followed by the collapse of skyrmions.

Magnetic solitons, such as skyrmions, antiskyrmions, and skyrmion bags, are viewed as promising candidates for future spintronic devices. Their electrical dynamics attract much interest from researchers, including electrical generation, deletion, and movement.[7, 8, 25, 56] In this work, we study the electrical response of the $Q = 0$ bag. By applying increasing pulsed currents (from $2.51 \times 10^{10}$ to $3.16 \times 10^{10}$ A m$^{-2}$) to the mixed state at $B = 117$ mT, the magnetic skyrmions annihilate gradually, forming a single $Q = 0$ bag at last (Figures 5a-d and Movie S2). Figure 5e shows the decrease of topological charge $Q$ as a function of current density $j$. Furthermore, at $B = 152$ mT, the $Q = 0$ bag shrinks and collapses to a single skyrmion soon under a pulsed current ($j \sim 2.73 \times 10^{10}$ A m$^{-2}$) with a pulse width of 20 ns (Figures 5f-i and Movie S3). The diameter of the $Q = 0$ bag as a function of pulse count in the deletion process is plotted in Figure 5j. At a fixed magnetic field, the $Q = 0$ bag is expected to have a single equilibrium configuration. However, it undergoes a discontinuous decrease in size due to the pinning effect, exhibiting no discernible movement upon injection of a pulsed current (Figure 5j). Throughout the pulse, multiple metastable states emerge, finally collapsing to a single magnetic skyrmion. This transformation is likely influenced by thermal effects or the spin transfer torque induced by the electrical current. The specific magnetic fields of 117 mT and 152 mT are necessary for the generation and deletion of skyrmion bags during current modulation. As illustrated in Figure S8, the stability of the $Q = 0$ bag is higher at lower magnetic fields, transforming into a single skyrmion within the higher field regime. Hence, a magnetic field of approximately 117 mT is advantageous for generating the $Q = 0$ bag electrically. Conversely, the $Q = 0$ bag is more susceptible to deletion at a higher magnetic field intensity of around 152 mT. Consequently, pulsed currents present a promising avenue for efficiently writing and deleting $Q = 0$ bag.

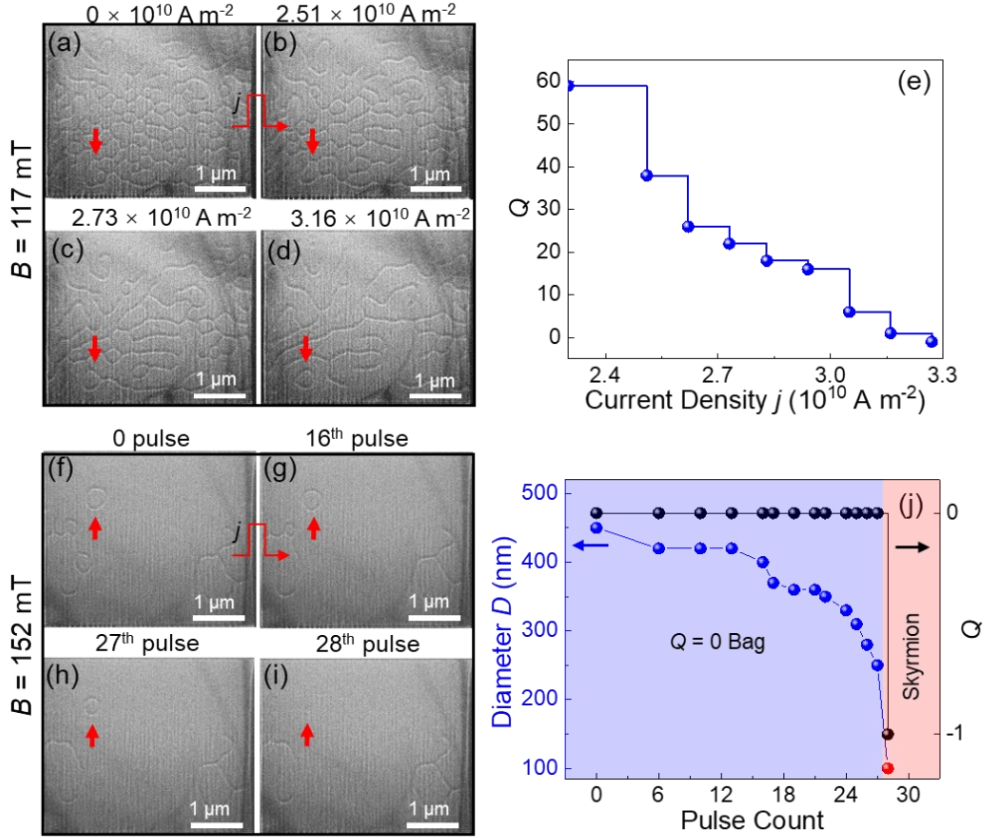

**Figure 5.** Current induced generation and deletion of $Q = 0$ bag at 295 K. (a)-(d) Current induced generation of $Q = 0$ bags with varied current density from $2.51 \times 10^{10}$ A m$^{-2}$ to $3.16 \times 10^{10}$ A m$^{-2}$ at $B = 117$ mT. (e) Topological charge $Q$ as a function of current density $j$. Each point represents a steady value of $Q$ after applying 20 pulses continuously. (f)-(i) Current induced deletion of a single $Q = 0$ bag with a current density of $2.73 \times 10^{10}$ A m$^{-2}$ at $B = 152$ mT. The pulse width is 20 ns. The defocused distance is −2.8 mm. $\alpha = -15°$. (j) The diameter of $Q = 0$ bag as a function of pulse count (left) and topological charge $Q$ as a function of pulse count (right). The red circle

and blue circle represent skyrmion and $Q = 0$ bag, respectively.

In summary, through the combination of the pulsed currents and magnetic fields, diverse Néel-twisted skyrmion bags are observed in a van der Waals magnet $Fe_{3-x}GaTe_2$ at and above room temperature. The thermal effect induced by pulsed currents promptly generates a mixed state comprising skyrmions and magnetic domains, which subsequently transitions into diverse skyrmion bags as well as double-nested skyrmion bags during the field-swapping process at room temperature. These skyrmion bags undergo a gradual reduction in inner skyrmions, eventually converting into single skyrmions under an increasingly perpendicular magnetic field. Furthermore, a thermal phase diagram of Néel-twisted skyrmion bags across room temperature regions is established, indicating a wide field and temperature range that a $Q = 0$ bag could exist stably. In addition, our exploration into the current-induced dynamical phase transformations of skyrmion bags has led to successful electrical generation and deletion of skyrmion bags. This study introduces a novel current-assistance method for generating Néel-twisted skyrmion bags in 2D van der Waals magnets, thereby advancing the development of $Q$-based spintronic devices.

**Methods**

**Sample preparation.** $Fe_{3-x}GaTe_2$ single crystals were grown by self-flux method with stoichiometric iron powders (99.99%), gallium chunks (99.99%), and tellurium powders (99.999%) mixed as a molar ratio of 1:1:2. The mixed raw materials were then sealed in a quartz tube and heated to 1000 °C at first. The high temperature was maintained for 24 hours to make the mixture react sufficiently. After the isothermal process, the temperature cooled down to 880 °C and then decreased slowly from 880 °C to 780 °C for 144 hours. Finally, keeping the quartz tube at 780 °C for 168 hours, thin $Fe_{3-x}GaTe_2$ single crystals were obtained after centrifugating the liquid mixture. The nanostructured $Fe_{3-x}GaTe_2$ lamellas were fabricated via a standard lift-out method from the bulk material by using a focused ion beam instrument (Helios Nanolab 600i, FEI).

**TEM Measurements.** Fresnel magnetic images were recorded by a Lorentz TEM instrument (Talos F200X, FEI) operating at 200 kV in Lorentz mode.[57] The perpendicular magnetic field was adjusted by manipulating the objective current, with pulsed currents being supplied by a voltage source (AVR-E3-B-PN-AC22, Avtech Electrosystems). A single tilt cryogenic holder (Gatan 613.6) is utilized to obtain the temperature range from 100 K to 370 K in-situ.

**Micromagnetic Simulations.** Micromagnetic Simulations were performed by using a GPU-accelerated micromagnetic simulation program, Mumax3.[58, 59] The exchange energy, Dzyaloshinskii-Moriya interaction (DMI) energy, Zeeman energy, and demagnetization energy were considered in the simulations. Magnetic parameters were set based on the $Fe_{3-x}GaTe_2$ material with $A_{ex}$ = 9 pJ/m, $K_u$ = 33.6 kJ/m$^3$, and saturation magnetization $M_s$ = 24.9 kA/m. The DMI interaction $D_{dmi}$ = 0.75 mJ m$^{-2}$ The cell size was set to 4 × 4 × 4 nm$^3$. The equilibrium spin configurations were obtained by using the conjugate-gradient method.

**Associated content**

The authors declare no competing financial interest.

**Supporting Information**

The Supporting Information is available free of charge on the ACS Publications website at http://pubs.acs.org.

**Movie S1:** Current induced generation of skyrmions with $j = 4.11 \times 10^{10}$ A m$^{-2}$ at $B = -8$ mT. The pulse width is 40 ns. $\alpha = +15°$. $T = 295$ K.

**Movie S2:** Current induced generation of $Q = 0$ bags with varied current density from $2.51 \times 10^{10}$ A m$^{-2}$ to $3.27 \times 10^{10}$ A m$^{-2}$ at $B = 117$ mT. The pulse width is 20 ns. $\alpha = -15°$. $T = 295$ K.

**Movie S3:** Current induced annihilation of a single $Q = 0$ bag with a current density of $2.73 \times 10^{10}$ A m$^{-2}$ at $B = 152$ mT. The pulse width is 20 ns. $\alpha = -15°$. $T = 295$ K.

**Figures S1 to S11:** Crystal characterization of Fe$_{3-x}$GaTe$_2$ single crystals. HRTEM image and corresponding SAED pattern of the Fe$_{3-x}$GaTe$_2$ (001) plane. Field-dependent magnetization curves at temperatures $T = 300, 320, 340, 360$, and 370 K, for $B$ is parallel to the *ab*-plane. Simulated Fresnel contrast of a single Néel-type skyrmion at different angles of $-15°$, $0°$, and $+15°$. Experimental Fresnel contrast of $Q = 0$ bag at different angles of $-27°$, $0°$, and $+27°$. Simulated field-driven process of the mixed state. Magnetic evolutions of the $Q = 0$ bags in the field-increasing and decreasing processes at 295 K. Magnetic field-dependent total energy curves during the field-increasing process from a single $Q = 0$ bag. Magnetic field-driven evolutions of the $Q = 0$ bag at 150 K. Field-increasing and decreasing process of the $Q = 0$ bags. Magnetic field-driven evolutions of the composite skyrmion bag with $Q = 0$ at 295 K.

**Author information**

Corresponding author

*(Y.W.) Email: wuyaodong@hfnu.edu.cn; *(Y.K.) Email: keyajiao@whut.edu.can;

*(J.T.) Email: jintang@ahu.edu.cn;

**Acknowledgments**

This work was supported by the National Key R&D Program of China, Grant No. 2022YFA1403603; the Natural Science Foundation of China, Grants No. 12422403, 12174396, 12104123, and U24A6001; the Anhui Provincial Natural Science Foundation, Grant No. 2308085Y32 and 2408085QA022; the Natural Science Project of Colleges and Universities in Anhui Province, Grants No. 2022AH030011 and 2024AH030046; Anhui Province Excellent Young Teacher Training Project, Grant No. YQZD2023067; the 2024 Project of GDRCYY (No. 217, Yaodong Wu); and the China Postdoctoral Science Foundation, Grant No. 2023M743543, and 2024M760006.

TOC Graphic

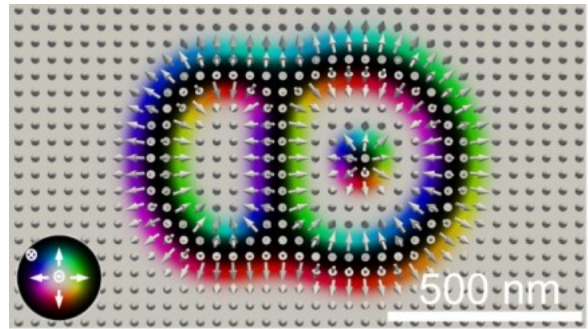

Double-nested skyrmion bags generated in $Fe_{3-x}GaTe_2$ at room temperature, by combing pulsed currents with the assistance of magnetic fields.